\documentclass{eas}
            \usepackage{graphicx}
\begin{document}

      \def\HI{H{\sc i} }
      \def\HII{H{\sc ii} }
      \def\Msun{\, {\rm M}_{\odot}}
      \def\OH{{12\-+log(O/H)} }
      \def\lNO{log(N/O) }

            \title{Star formation regulation, gas cycles and the chemical 
      evolution 
            of dwarf irregular galaxies} 
            \runningtitle{Star formation regulation, gas cycles and chemical 
                          evolution}
            \author{Simone Recchi$^{1, }$}
            \address{Institute of Astronomy, Vienna University, 
            T\"urkenschanzstrasse 17, 1180 Vienna, Austria}
            \address{INAF - Osservatorio Astronomico di Trieste, 
            Via G.B. Tiepolo 11, 34131 Trieste, Italy}
            \author{Gerhard Hensler$^1$}
            \begin{abstract}

      Due to their low gravitational energies, dwarf galaxies are greatly
      exposed to energetical influences from internal and external sources.
      By means of chemodynamical models we show that their star formation is
      inherently self-regulated, that peculiar abundance ratios can only
      be achieved assuming different star-formation episodes and that
      evaporation of interstellar clouds embedded in a hot phase can lead to
      a fast mixing of the interstellar gas. Metal-enriched hot outflows
      can accrete onto infalling clouds by means of condensation leading
      to a large range of timescales for the self-enrichment of the ISM from
      local scales within a few tens of Myr up to a few Gyr for the
      large-range circulation. Infall of clouds is also required to explain
      abundance ratios of metal-poor galaxies at evolved stages because it
      reduces the metallicity altering only marginally the abundance ratios.

            \end{abstract}
            \maketitle
            \section{Introduction}

      Gas-rich dwarf galaxies are commonly thought to be characterized
      by a {\it bursting} star formation (SF), namely, short episodes
      of intense SF rates separated by long periods of quiescence. An
      alternative scenario is the {\it gasping} one, namely, long
      episodes of moderate SF rates separated by short periods of SF
      suppression. The presence of stars of intermediate age is
      apparently ubiquitous (e.g. Kunth \& \"Ostlin \cite{ko00}) ,
      therefore these object are more evolved than previously
      thought. The chemical evolution of galaxies is strongly affected
      by the choice of the SF regime and by environmental effects, in
      particular gas infall. In this contribution, we will focus on
      the evolution of abundance ratios under different astrophysical
      assumptions and on the cooling and mixing timescales of freshly
      produced metals, a complex, very controversial and still
      unsolved problem (see e.g. Roy \& Kunth \cite{rk95};
      Tenorio-Tagle \cite{tt96}; de Avillez \& Mac Low \cite{dm02}
      among others).

      \section{Single gas-phase chemodynamical model of gas-rich dwarf galaxies}
            \label{model}


            \subsection{The chemical and dynamical evolution after 
                        instantaneous starbursts}
            \label{results_burst}

      An intense SF episode of short duration distributes the
      energetic and chemical feedback from massive stars in a few tens
      of Myr. If this is the first significant episode of SF, a large
      fraction of freshly produced metals can cool and mix with the
      surrounding ISM in a timescale of the order of a few 10$^7$
      yr. This is shown by {\it single gas-phase} chemodynamical
      models of dwarf irregular galaxies (dIrrs) (Recchi et
      al. \cite{recc01}; \cite{recc02}) and occurs essentially for two
      reasons. On the one hand, the thermalization efficiency of SNe
      exploding in a cold and dense medium is very small. A long time
      elapses before the expansion velocity of the supernova remnant
      (SNR) decreases to become equal to the local sound speed (time
      at which the cavity of the SNR and the external medium get in
      casual connection) and in this interval radiative losses reduce
      considerably the thermal budget of the SNR. On the other hand, a
      supershell which is not fast enough to create soon a break-out
      of the galaxy is subject to the growth of Kelvin-Helmholtz and
      Rayleigh-Taylor instabilities and to the development of large
      eddies.  Thermal conduction has also time to spread the
      conduction front separating the cavity from the supershell.
      These phenomena favor the cooling of the metal-rich material
      inside the cavity and its mixing with the external unprocessed
      ISM. This implies that a significant fraction of the freshly
      produced chemical elements leaves quite soon (in $\sim$ 10--20
      Myr) the hot phase (where it is undetectable in the optical) and
      contributes to the observed metallicity of the galaxy, which
      therefore increases quite rapidly in the early evolution of the
      object.

            \subsection{Gasping or continuous star formation and consequences 
      for the chemical evolution of galaxies}
            \label{results_gasp}

      This non-bursting SF scenario is characterized by a much
      smoother energy release and therefore also the thermodynamical
      properties of the ISM change gently.  Eventually a galactic wind
      arises after timescales of 100 Myr or more and at this point the
      accumulated energy of several generations of stars has carved a
      large cavity. Every subsequent generation of stars releases its
      energy in a hotter and more tenuous medium.  The cooling
      timescale of this gas is much larger and also the occurrence of
      eddies or thermal instabilities is more unlikely.  Therefore it
      is much more complicated for the gas released by late episodes
      of SF to cool down and mix with surrounding cold gas.  In
      figs. 6 and 7 of Recchi et al. (\cite{recc06}), this effect is
      visualized for a model reproducing NGC~1569, one of the best
      studied proto-typical starburst dIrrs.  It is clearly visible
      that the fraction of cooled ejecta from the last episode of SF
      is very large for a bursting model whereas it is negligible
      compared to the hot ejecta in a gasping model.

            \begin{figure}
              \label{fig:on}
            \vspace{-1cm}
              \hspace{0.6cm} \includegraphics[height=0.8\columnwidth]
            {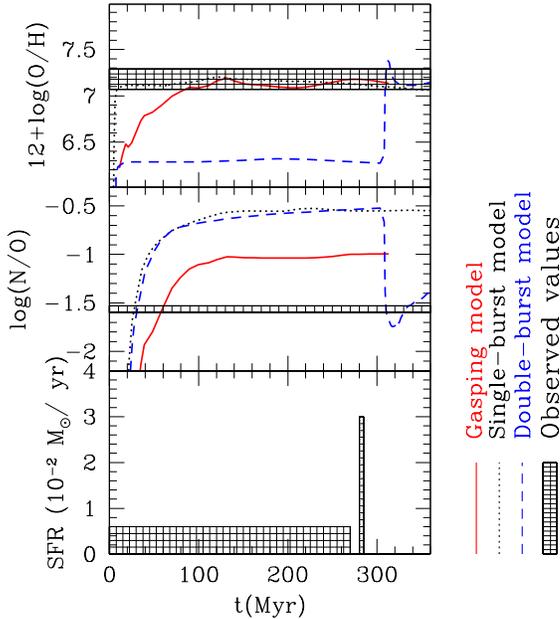} 
            \vspace{-1.2cm}

            \caption{ Evolution of \OH (top panel) and \lNO (central panel)
            for models reproducing I Zw 18 with a gasping SF (solid line),
            with one (dotted line) or two (dashed line) instantaneous
            bursts. The SF history of the gasping model is represented by
            the shaded area in the bottom panel.} \end{figure}

More importantly, these different behaviors change the chemical
pattern of a galaxy substantially. For models reproducing the main
properties of I Zw 18, one of the most metal-poor galaxies locally
known, we have compared in Fig.~1 the evolution of O and N/O for a
gasping model (whose SF history is sketched in the bottom panel) and
models with one or two instantaneous bursts of SF (see Recchi et
al. \cite{recc04}).  For this, only the metallicity of gas with
temperatures below $2 \cdot 10^4$ K is considered. As expected, due to
the fast cooling timescales of the ejecta, the bursting models are
characterized by a fast change of the abundance ratios in short
timescales. Gasping models show instead a continuous increase of the
metallicity for about 120 Myr. Thereafter, the cooling timescale of
the ejecta becomes so large that they cannot contribute anymore to the
global chemical enrichment of the galaxy, therefore the metallicity
and the abundance ratios do not change appreciably from this moment
on. We notice in particular that, after the last burst occurring at
280 Myr, no increase of O or decrease of N/O (as it would be expected)
is observed. In fact, almost all the ejecta produced by this burst are
either carried out of the galaxy by a galactic wind or are released
into a too hot phase, where they remain due to its very long cooling
timescale.

      \section{Large-scale matter cycle in multi-phase chemodynamical models}

      dIrrs with high SF rates are often surrounded by large \HI
      reservoirs with decoupled dynamics (e.g.\ NGC 4449, I Zw 18) or
      are suffering a collision with large intergalactic \HI clouds
      (e.g.\ He~2-10, II~Zw~40). This leads to the plausible
      assumption that SF is triggered and even enhanced to a burst by
      gas infall if the infalling \HI clouds replenish the consumed
      gas on a timescale shorter than that of self-regulating energy
      release by massive stars.  Gas infall is observationally
      manifested for NGC~1569 where, from an extended and tidally
      disrupted \HI cloud complex (Stil \& Israel 2003), a series of
      gas clumps falls in towards NGC~1569 (M\"uhle et al.\
      2003). This striking infall scenario leads to several
      consequences that can be studied in models and interpreted from
      observations. In particular, gas infall affects the outflow
      cycle of hot metal-enriched gas and the chemical abundance
      ratios (assuming that the infalling clouds have metallicities
      smaller than those in a galaxy).

      Because of the coexistence of cold clouds enveloped by hot
      metal-rich gas and due to turbulence and the fragmentation of
      superbubble shells, the chemical elements freshly released by
      SNeII are mixed into the cool gas within the SF environment (see
      also Sect.~\ref{results_burst}). If the hot gas is able to
      evaporate the cold clouds and due to its overpressure a galactic
      mass-loaded outflow occurs (Hensler et al. 1999). Since these
      clouds contain elements from intermediate-mass stars of older
      stellar populations, e.g.\ carbon and nitrogen, the N/O ratio
      should represent this mixing effect. The cooling of hot gas
      enables its condensation onto infalling clouds and leads to
      their pollution with the elemental mix.  The abundance patterns,
      however, become apparent only after the formation of \HII
      regions.

      The analyses of this matter cycle in a {\it multi-phase}
      chemodynamical dIrr model of $10^9 \Msun$ baryonic mass in a
      $10^{10} \Msun$ dark matter halo (Hensler 2001) show that $\sim$
      25\% of the metals produced in massive stars remain within the
      SF sites and lead to a local self-enrichment within 1 kpc on
      typical timescales in the range of 10 Myr. The remaining 3/4 of
      produced SNII metals are carried away from the SF region by the
      superbubble expansion.  Since the temperature of the hot gas
      decreases with its expansion, i.e.  is lower at larger
      distances, the fraction of condensed hot gas and, by this, of
      the metal deposition in clouds increases outwards up to a
      distance of 15 kpc from the dIrr's center.

      Although the metals are hardly directly expelled from a dIrr by
      the galactic wind but incorporated into infalling clouds, the
      circulation timescale for the return of metals originating in
      the galaxy can last from 1 Gyr at 3 kpc to 10 Gyrs from above 10
      kpc (Hensler 2001). If one takes into account that widely
      distributed metal-enriched hot gas in dIrrs can be stripped off
      by the intergalactic medium or by tidal effects, still 50\% of
      the metals from SNeII are retained in this model and transferred
      to the cool gas within a distance of not more than 8 kpc.  As a
      result, analytical studies and hydrodynamical models that
      investigate the expansion of hot SNII gas alone as tracer of the
      metal dispersal, overestimate the total metal loss from the
      galaxy if small-scale mixing effects are neglected.

      \section{The effect of gas infall on the element abundances}

      In the N/O-O/H diagram dIrrs form a cloud with an appreciable
      amount of scatter around \lNO = -1.5 and over a range of \OH
      between 7 and 8.5. This regime is usually passed by evolving
      galaxies within their very early stages of evolution although
      the path can vary due to the SF timescales (Henry et
      al. 2000). Since the vast majority of dIrrs contains old stellar
      populations, the problem exists how to reach these low abundance
      values from an evolved state that is reached on the secondary
      production track of N. Under the assumption that dIrrs are young
      systems several authors allowed for starburst-driven galactic
      winds with selective element depletion, while others presented
      models that follow the assumption of abundance self-enrichment
      of the observed \HII regions. For a comprehensive discussion see
      Hensler et al. (2004).

      \begin{figure}
            \vspace{-1cm}
              \includegraphics[width=8cm,angle=-90]
            {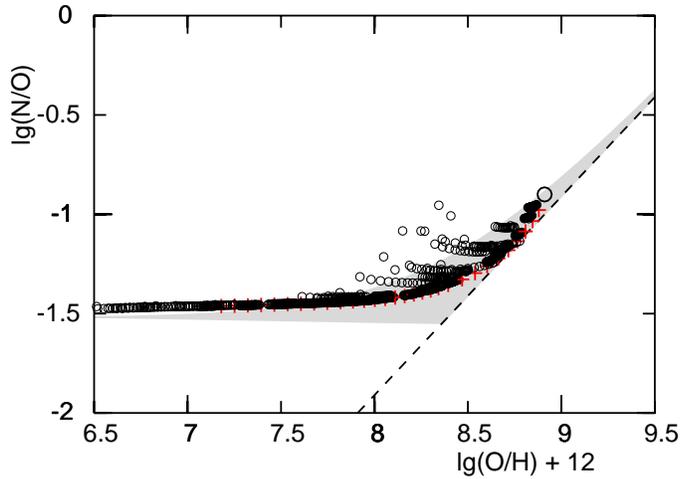} 
              \label{infall}
      \vspace{-0.5cm}
            \caption{Location of dIrrs models with infall and starbursts
             in the \lNO - \OH diagram. For details see text. }
            \end{figure}

      On the reasonable basis of the infall scenario of pristine gas
      K\"oppen \& Hensler (2005) explored the evolutionary path in the
      \lNO - \OH phase space when starting from an evolved state with
      a metal enrichment by former stellar populations. The main issue
      is that the path forms a loop at first to lower O/H values and
      back to the starting point via lower N/O ratios. The extent of
      the loop is related to the mass ratio of infalling to existing
      gas in the SF site. Since galaxies with smaller masses remain
      underdeveloped at lower \lNO - \OH values than more massive
      ones, the loop sizes are larger for low-mass galaxies.  As a
      result, a sampling of all possible infall models therefore
      reproduces well the characteristic observed distribution (Henry
      \& Worthey 1999) which cover an almost triangular regime with
      larger extension at lower N/O. Taking additionally a reasonable
      infall-triggered starburst into account the distribution
      (Fig.~2) changes not much in comparison with burst-less models
      (see figs. 18 and 20 in K\"oppen \& Hensler 2005). In a very
      recent study Knauth et al. (2006) have also invoked gas infall
      as an explanation for the local interstellar N/O abundances.

      %
      %

            \section{Concluding remarks}
            \label{conclusions}

            The main results of our work can be summarized as follows:

            \begin{itemize}
       \vspace{-0.3cm}

            \item short episodes of SF enrich the ISM in a timescale
            of a few tens of Myr.\vspace{-0.2cm}

            \item Long-lasting episodes of SF enrich gradually the ISM
            in a longer timescale. Any further episode of SF does not
            leave an appreciable imprint on the chemical evolution. In
            fact, the metals produced by these SF episodes are either
            directly carried out of the SF region, or they are
            released in a too hot medium and they do not have the
            chance to pollute the surrounding ISM.\vspace{-0.2cm}

            \item Multiphase chemodynamical models of dIrr galaxies
            show that $\sim$ 25\% of the stellar ejecta mix locally
            (within 1 kpc) on typical timescales of the order of 10
            Myr, whereas the remaining metals undergo a longer cycle,
            via condensation onto clouds on timescales larger than 1
            Gyr.\vspace{-0.2cm}

            \item The infall of metal-poor clouds changes the chemical
            evolution of dIrrs. In particular it forms loops in the
            O/H vs. N/O diagram, whose extension is related to the
            mass ratio of infalling to existing gas in the SF
            cloud. These loops account well for the observed N/O of
            metal-poor dwarf galaxies.

            \end{itemize}

      \acknowledgements

            Decisive contributions to the presented work are acknowledged
            from our collaborations with L. Angeretti, A. D'Ercole,
            J. K\"oppen, F. Matteucci, A. Rieschick, Ch. Theis, and M. Tosi.
            This work is partly supported (S.R.) by the Deutsche
            Forschungsgemeinschaft under grant HE1487/28.

            \endacknowledgements


            \end{document}